\documentclass[12pt]{article}

\usepackage{amsmath,amssymb,amsthm,graphicx}

\newcommand{\ed}{\end{document}}
\newcommand{\be}{\begin{equation}}
\newcommand{\ee}{\end{equation}}

\begin{document}
\begin{center}
\large{\textbf{CONSTRAINED DYNAMICS OF AN ANOMALOUS $(g\neq 2)$ RELATIVISTIC
SPINNING PARTICLE IN ELECTROMAGNETIC BACKGROUND}}\\
\end{center}

\begin{center}
A. Berard$^{c, }$\footnote{E-mail:aberard001@noos.fr},
Subir Ghosh$^{b,}$\footnote{E-mail: sghosh@isical.ac.in},
Y. Grandati$^{c, }$\footnote{E-mail:grandati@yahoo.fr },
\\
H. Mohrbach$^{c, }$\footnote{E-mail:Herve.mohrbach@univ-metz.fr}, Probir Pal
$^{a,}$,
\\
$^a$     S. N. Bose National Centre for Basic Sciences,
Block-JD, Sector-III, Salt Lake, Kolkata 700 098, India\\
$^b$Physics and Applied Mathematics Unit, Indian Statistical
Institute\\
203 B. T. Road, Kolkata 700108, India \\
$^c$Laboratoire de Physique Mol ́culaire et des Collisions, ICPMB, IF CNRS
2843,\\
 Universite ́ Paul Verlaine,
Institut de Physique,\\
 Bd Arago, 57078 Metz, Cedex 3, France
\end{center}\vspace{1cm}

\begin{center}
\textbf{Abstract}
\end{center}
In this paper we have considered the dynamics of an anomalous ($g\neq 2$)
charged relativistic spinning particle
in the presence of an external electromagnetic field. The constraint analysis is
done and the complete set of Dirac
brackets are provided that generate the canonical Lorentz algebra and dynamics
through Hamiltonian equations of
motion. The spin-induced effective curvature of 
spacetime and its possible connection with Analogue Gravity models are commented
upon.
\vspace{.5cm}

   \newpage
The symplectic structure (which, in the special case of a point particle
lagrangian in a first order
formalism,  is essentially the terms containing time
derivatives) completely dictates the kinematics of the
particle as it generates the phase space algebra. This
in turn determines the correct structure of Lorentz generators (or angular
momenta) that will satisfy
the Lorentz algebra. The mass-energy dispersion relation is also induced from
the symplectic structure. Through
Hamilton's equations the dynamics is determined once a potential energy term is
prescribed. The canonical symplectic
structure, $\sim~p^i\dot q_i$ ($p^i,q_i$ being the momenta and coordinates
respectively) yields the canonical set of Poisson Brackets (PB)
$\{q_i,q_j\}=\{p^i,p^j\}=0,~\{q_i,p^j\}=\delta_i^j$. But more complicated form
of symplectic structures
$\sim~p^i A_i^j(p,q)\dot q_j$, in general yields more complicated brackets
depending up on the specific form of $A_i^j(p,q)$.  In general
the latter is referred as Non-Commutative (NC) phase space since  additional
terms in the canonical PB
 structure appear. NC spacetime is a special case where $\{q_i,q_j\}\ne 0$.
There are two
related ways to derive the algebra from a given
symplectic structure: the Dirac constraint analysis formalism \cite{dir} or
Faddeev-Jackiw symplectic formalism \cite{fadd}, both
obviously leading to the same results.  In the present work we will exploit the
Dirac formalism. The symplectic structure
induces constraints that determine the generalization of PB, known
as Dirac Brackets (DB), via a systematic
procedure. In a particular problem it is possible that one can fix the
constraints in
an intuitive way from the physics of the
problem. In that case one does not need the explicit form of the symplectic
structure and can directly compute
the DB and subsequently obtain the particle kinematics and dynamics.
This is the case for a charged relativistic
spinning particle in the presence of an external electromagnetic field, to be
discussed below. As we will mention at the end of the
paper the present work can have
interesting consequences in the topically active areas of Analogue Gravity
models \cite{riv}.

Before proceeding to the main topic let us mention some earlier works to get the
perspective of the present work. The
NC phase space - relativistic spinning particle connection in $3+1$-dimensions
appeared explicitly first in \cite{regge}. Later
similar types of spinning particle models \cite{jac,sg,ch1} became popular in
the context of Anyons, excitations of arbitrary spin in
$2+1$-dimensions \cite{wil}. We will closely follow the work of Chaichian,
Gonzalez Felipe and D.Louis Martinez \cite{ch2}. We aim to
construct the NC phase space for a $3+1$-dimensional classical spinning particle
of charge $e$ and an anomalous gyromagnetic ratio $g\neq 2$ in
the presence of a constant and weak electromagnetic field. The resulting
Hamiltonian model in our case
turns out to be structurally different from that of \cite{ch2} for reasons that
will be discussed as we proceed.
Furthermore, we will provide a detailed constraint
analysis, compute DBs and subsequently apply them in revealing the dynamics of
the degrees of freedom.
 Our results
will be
valid to first non-trivial order in the electromagnetic field.

With this brief introduction we now proceed to the main body of our
work.  The  phase space degrees of freedom  with the following set of
coordinate and spin variables, $x_\mu,~p_\mu$ and
$n_\mu,~p^{(n)}_\mu$ respectively with $\mu =0,1,2,3$ and our
metric is $\eta_{\mu\nu},~\eta_{00}=-\eta_{ii}=1$. The above canonically
conjugate pairs
are independent with the PB,
\be
\left\{ {{x}_{\mu \,,}}\,{{p}_{\nu }} \right\}\,=-\eta _{\mu \nu },
\,\,\{\,{{x}_{\mu \,\,,}}\,
{{x}_{\nu \,}}\}\,=\,\{\,{{p}_{\mu \,,\,}}{{p}_{\nu }}\,\}\,=\,0 ,
\label{pbx}
\ee
\be
\left\{ {{n}_{\mu \,,}}\,{{p}^{(n)}_{\nu }} \right\}\,=-\eta _{\mu \nu },
\,\{{{n}_{\mu \,}},\,{{n}_{\nu }}\,\}\,=\,\{\,{{p}^{(n)}}_{\mu
}\,, \,{{p}^{(n)}}_{\nu }\,\}\,=\,0. \label{pbx1}
\ee
The $n_\mu ,p^{(n)}_\mu $ sector will describe the spin. All cross brackets
between
$n_\mu ,p^{(n)}_\mu $ and $x_\mu ,p_\mu $ vanish.

The particle moves in a constant electromagnetic field
$F_{\mu\nu}$ given in terms of the potential by $A_\mu =
-\frac{1}{2}F_{\mu\nu}x^\nu $. Sometimes it is convenient to use
the variable ${{\pi }_{\mu }}=\,{{p}_{\mu }}-\,e\,{{A}_{\mu }}$
with the canonical PB
\be\{{{\pi }_{\mu ,}}{{\pi }_{\nu }}\}=-e{{F}_{\mu \nu }}\label{ph}
\ee
We also consider the total angular momentum operator $J_{\mu\nu}$
consisting of a rotation part $l_{\mu\nu}$ and a spin part
$s_{\mu\nu}$,
\be J_{\mu\nu}= l_{\mu\nu}+s_{\mu\nu}= (x_\mu p_\nu -x_\nu p_\mu )+(n_\mu
{p}^{(n)}_\nu -n_\nu {p}^{(n)}_\mu) ,
 \label{npn}
\ee
obeying the conventional Lorentz algebra,
\be\,
\{J_{\mu\nu},J_{\alpha\beta}\} ={{J}_{\mu \beta }}{{g}_{\nu \alpha
}}+{{J}_{\nu \alpha }}{{g}_{\mu \beta }}+{{J}_{\alpha \mu }}{{g}_{\nu \beta }}
+{{J}_{\beta \nu }}{{g}_{\mu \alpha }}.\label{4}
\ee

In a relativistic formulation of spinning particle one needs to
introduce constraints such that the spin tensor reduces to a three
dimensional vector in the particle rest frame
\cite{regge,ch1,ch2,jac,sg}. This is the Hamiltonian analogue of
the well known Frenkel condition $$(s_{\mu\nu }\dot x^\nu
)/d\tau =0.$$
  The constraints are the following \cite{ch1,ch2} (see also \cite{regge,sg}):
\be\Phi_1\equiv (\pi {{p}^{(n)}})\approx 0;\,\,\chi _2 \equiv (\pi \,n\,)
\approx 0. \label{c}
\ee
We use a shorthand notation $A^\mu B_\mu =(AB).$ This pair of
Hamiltonian constraint is Second Class (in the sense of Dirac
constraint analysis \cite{dir}) that is they do not commute at the
PB level,
\be\{{{\Phi }_{1}},{{\Phi }_{2}}\}={{\pi }^{2}}+\frac{e}{2}\,
{{F}_{\mu \nu }}{{s}^{\mu \nu }}.\label{e2}
\ee
Hence we are obliged to replace the PB by DB as
defined below for two generic dynamical variables $A,B$,
\be\{A,B\}_{DB}=\{A,B\}_{PB}-\{A,\Phi _i\}_{PB}({\{\Phi _i,\Phi
_j\}_{PB})}^{-1}\{\Phi _,B\}_{PB}. \label{db}
\ee
From here on all the brackets are DB and hence we drop
the subscript $DB$ in the bracket. The constraints can also be
treated as strong relations, $\Phi_1=\Phi _2 =0$. The constraint
matrix and  its inverse are given by,
$$\{\Phi _i,\Phi _j\}_{PB}=\,\,\,\,\left( \,\begin{matrix}
   0 & {{\pi }^{2}}+\frac{e\,}{2}\,\,{{F}_{\mu \nu }}{{s}^{\mu \nu }}  \\
   -{{\pi }^{2}}-\frac{e}{2}\,{{F}_{\mu \nu }}{{s}^{\mu \nu }} & 0  \\
\end{matrix} \right)\,
$$$$\{\Phi _i,\Phi _j\}_{PB}^{-1}=\,\,\left( \begin{matrix}
   0 & \frac{-1}{{{\pi }^{2}}+\,\,\,\frac{e}{2}{{F}_{\mu \nu }}{{s}^{\mu \nu }}}
 \\
   \frac{1}{{{\pi }^{2}}+\frac{e}{2}{{F}_{\mu \nu }}{{s}^{\mu \nu }}} & 0  \\
\end{matrix} \right).$$
Exploiting the definition of DB (\ref{db}) we compute the latter,

\be\{{{x}_{\mu ,}}{{x}_{\nu }}\}=\frac{{{s}_{\mu \nu }}}{{{\pi }^{2}}
+\frac{e}{2}{{F}_{ \alpha\beta}}{{s}^{ \alpha\beta}}},\{{{x}_{\mu
}},\,{{\pi }_{\nu }}\}=-{{\eta }_{\mu \nu }}-\,\frac{e{{s}_{\mu}^{~
\alpha }} {{F}_{\alpha \nu }}}{{{\pi }^{2}}\,+\frac{e}{2}{{F}_{ \alpha\beta}}
{{s}^{ \alpha\beta}}},\{{{x}_{\mu }},{{n}_{\nu }}\}
=\frac{{{n}_{\mu }}{{\pi }_{\nu }}}{{{\pi }^{2}}\,
+\frac{e}{2}{{F}_{\alpha\beta }}{{s}^{\alpha\beta }}}\label{m}
\ee

\be\{{{x}_{\mu }},{{p}_{\nu }}^{(n)}\}=\,\frac{{{p}^{(n)}}_{\mu }{{\pi }_{\nu
}}}{{{\pi }^{2}}+\frac{e}{2}{{F}_{\alpha\beta }}{{s}^{\alpha\beta
}}}\,, \{{{\pi }_{\mu }},{{\pi }_{\nu }}\}=-e{{F}_{\mu
\nu }}\,,\{{{n}_{\mu }},{{p}^{(n)}}_{\nu }\}=-{{g}_{\mu \nu }}
+\frac{{{\pi }_{\mu }}{{\pi }_{\nu }}}{{{\pi }^{2}}
+\frac{e}{2}{{F}_{ \alpha\beta}}{{s}^{\alpha\beta }}} \label{1}
\ee

\be
\{{{p}^{(n)}}_\mu,{{\pi }_{\nu }}\}=\frac{e{{\pi }_{\mu }}
{{p}^{(n)}}^{\alpha }{{F}_{\alpha \nu }}}{{{\pi }^{2}}
+\frac{e}{2}{{F}_{\alpha\beta }}{{s}^{\alpha\beta}}}
,\{{{n}_{\mu }},{{\pi }_{\nu }}\} =\frac{e{{\pi
}_{\mu }}{{n}^{\alpha }}{{F}_{\alpha \nu }}}{{{\pi }^{2}}
+\frac{e}{2}{{F}_{\mu \nu }}{{s}_{\mu \nu }}}\,$$$$
\{{{n}_{\mu }},{{n}_{\nu }}\}=\{{{p}_{\mu }},
{{p}_{\nu }}\}=\{{{p}^{(n)}}_{\mu },{{p}^{(n)}}_{\nu }\}
=0\,.\label{2}
\ee
This shows that the constraints induce a NC spacetime. Notice the with respect
to the DBs the spin and coordinate sectors have become mixed up. The above set
of DBs are exact.

It is very important from our perspective to note that using the
DBs (\ref{m},\ref{1},\ref{2}) the angular
momentum $J_{\mu\nu}$ having the same canonical definition as
given in (\ref{npn}) satisfies the {\it{undeformed}} Lorentz algebra
(\ref{4}) to $O(e)$. It is now straightforward to construct the
Hamiltonian constraint (or equivalently the mass shell condition):
{\it{we keep all terms that are Lorentz invariant, that is commutes in the sense
of DB with
$J_{\mu\nu}$, to $O(e)$}}.
This is our guiding principle. Explicitly the Hamiltonian is,
\be
H= \frac{1}{m} [\frac{\pi
^2}{2}+\frac{eg}{2}F_{\alpha\beta}J^{\alpha\beta}+\frac{e(g-2)}{m^2}F_{
\alpha\beta}J^\beta
_{~\gamma }p^\alpha p^\gamma ]+\nu n_\alpha n^\alpha +\lambda
p^{(n)\alpha}p^{(n)}_\alpha +\sigma n^\alpha p^{(n)}_\alpha .
\label{ham}
\ee
In the above $\nu,\lambda,\sigma $ are $c$-number parameters.
Note that we have kept the same parametrization of constants $e,g$ as
that of \cite{ch2} but the terms are  distinct from
those of \cite{ch2}. $g$ is identified as the gyromagnetic ratio
and for $g=2$ one recovers the conventional case with the last
term in $H$ being absent. The last term is a manifestation of
anomalous value of $g\neq 2$.

It is worthwhile to compare and contrast our formalism and results with other
earlier works in the same area: in 
particular that of van Holten \cite{jan} and Chaichian et.al. \cite{ch2}. At the
outset we point out that
our formalism is more mathematical or algebraic in nature since we base our
construction of the cherished Hamltonian operator, (that generates the dynamics via Dirac Brackets
computed here), purely by 
demanding that it be Lorentz invariant, (that is commutes with the Lorentz
generators again in the sense of Dirac
Brackets). On the other hand the abovementioned works are more intuitive and are
principally based on known
(or expected)  physical
behaviour of a classical charged spinning particle in external field. Hence our
setup is more general and but can
be reconciled with the physically oriented systems discussed in \cite{jan,ch2}.
In explicit terms one
important difference between our Hamiltonian and the previous ones in
\cite{jan,ch2} for $g=2$
 is that in our case the coupling term is
$eJ_{\mu\nu}F^{\mu\nu}=e(l_{\mu\nu}+s_{\mu\nu})F^{\mu\nu}$ 
whereas in the other it is $es_{\mu\nu}F^{\mu\nu}$. In
our case we have ensured Lorentz invariance. Furthermore in \cite{jan} there is
a prediction of a 
novel form of relativistic time dilation effect even for a particle at rest.
Similar effect will appear hare as well but with some
additional complication coming from  the extra term $el_{\mu\nu}F^{\mu\nu}$ in
our Hamiltonian. Again in \cite{jan}
the author has constructed a parallel setup in terms of anticommuting degrees of
freedom to simulate
spin dynamics of the corresponding quantum system where an interacting
Klein-Gordon form of equation for the
spinning particle is advocated with the canonical identification of $
p_j=-i(\partial/\partial x_j)$. In principle
this can be carried through in our formalism but more work is involved since in
our basic Dirac Bracket framework
$p_i,x_j$ and $x_i,x_j$ do not obey canonical algebra. It will be necessary in
our framework to find a Darboux
map to a set of canonical variable with which one can proceed.

The $g\ne 2$ was
studied in detail in \cite{ch2}. One can now understand the difference in the expressions of the
Hamiltonian between our result (\ref{ham}) and \cite{ch2}. In
\cite{ch2} the authors have concentrated on keeping the
Hamiltonian constraint First Class (in the Dirac sense \cite{dir})
that is commuting with all other constraints whereas we have
ensured that the Hamiltonian is Lorentz invariant that is it
commutes with the Lorentz
 generators $J_{\mu\nu}$ in the DB sense. The
dynamics given in \cite{ch2} (see equations 25-26), reduced to $O(e)$,
essentially matches similar equations
(\ref{e3}-\ref{dots}) in our model with the obvious differences coming from the
extra terms present in our Hamiltonian.

Exploiting the DBs (\ref{m}, \ref{1},\ref{2}) once again we can evaluate the
dynamics by taking commutators between each dynamical variable and
$H$ in (\ref{ham}). We give the result for $\dot \pi _\mu $ in detail (keeping
terms of $O(e)$ throughout the rest of 
the paper):
\be
\{\pi _\mu ,\frac{\pi^2}{2m}\}=-\frac{e}{m}F_{\mu\nu}p^\nu
\label{e1}
\ee

\be
\{\pi _\mu
,\frac{eg}{2m}F_{\alpha\beta}J^{\alpha\beta}\}=\frac{eg}{m}F_{\mu\nu}p^\nu
\label{e21}
\ee

\be
\{\pi _\mu ,\frac{e(g-2)}{m^3}F_{\alpha\beta}J^\beta
_\gamma\pi^\alpha \pi^\gamma \}=-\frac{e(g-2)}{m}F_{\mu\nu}p^\nu
\label{e3}
\ee
Combining the above we find,
\be
\dot \pi_\mu =\{\pi_\mu ,H\}=\frac{e}{m}F_{\mu\nu}p^\nu.
\label{eqp}
\ee
Thus from (\ref{eqp}) it is clear that  there is no deviation from the
Lorentz force to $O(e)$ and the result is indendent of $g$.

 The time derivatives of rest of
the degrees of freedom are,

\be
\dot x_\mu =-\frac{1}{m}(\pi_\mu
+e\frac{1}{m^2}s_{\mu\nu}F^{\nu\rho}p_\rho
+e\frac{(g-2)}{m^2}(F_{\mu\nu}J^{\nu\rho}+J_{\mu\nu}F^{\nu\rho})p_\rho
) \label{eqx}
\ee

\be
\dot n_\mu =\frac{eg}{m}F_{\mu\nu}n^\nu
+\frac{eg}{m^2}(g-1)F_{\nu\rho}n^\nu p^\rho p_\mu -\sigma n_\mu
, \label{n
}
\ee

\be
\dot p^{(n)}_\mu =\frac{eg}{m}F_{\mu\nu}p^{(n)\nu }
+\frac{e(g-1)}{m^3}F_{\nu\rho}p^{(n)^\nu }p^\rho p_\mu +\sigma
p^{(n)}_\mu ,\label{eqpn}
\ee
From the above relations it is straightforward to compute
\be
\dot s_{\mu\nu}= \frac{2e}{m^3}(s_{\mu\kappa}\pi_\nu
-s_{\nu\kappa}p _\mu )p_\lambda
F^{\kappa\lambda}-\frac{eg}{m}(s_{\mu\alpha}g_{\nu\beta}-s_{\nu\alpha}g_{
\mu\beta})F^{\alpha\beta}$$$$
+\frac{e(g-2)}{m^3}(s_{\beta\mu}p_\nu -s_{\beta\nu}p_\mu
)p_\alpha F^{\alpha\beta} . \label{s}
\ee
We define the Pauli-Lubanski vector as
\be
s_\mu =\epsilon _{\mu\nu\alpha\beta}s^{\nu\alpha}\pi^\beta
 \label{pl}
\ee and compute \be \dot s_\mu =\frac{e}{m}F_{\mu\nu}s^\nu
-\frac{e}{2m^3}(g-2)p_\mu F_{\alpha\beta}p^\alpha s^\beta
.\label{dots} \ee It is very important to check that the Frenkel
condition is maintained as this ensures that the relativistic spin
tensor $s_{\mu\nu}$ reduces to the spin vector in the particle
rest frame. In our generalized system we find \be d(s_{\mu\nu}\dot
x^\nu )/d\tau =\frac{e\sigma
(g-2)}{m^2}s_{\mu\nu}(F^{\nu\alpha}J_{\alpha\beta}p^\beta
+J^{\nu\alpha}F_{\alpha\beta}p^\beta ). \label{w} \ee This
relation demands that we restrict our model to $ \sigma =0$ in
(\ref{ham}) so that the Frenkel condition is time invariant. This ensures
further that \be \ddot s_\mu =0.
\label{dds} \ee 

 In the present work we have studied the spinning particle
model in a Hamiltonian framework. Hamiltonian analogues of the
Frenkel condition appear as constraints and yields a
non-commutative spacetime through the Dirac Brackets. The latter
are essential for a proper quantum treatment of the model. This is
beyond the scope of the present paper as the operatorial nature of
the Dirac Brackets make the quantization non-trivial. However
there is another area where the present work can have interesting
consequences: Analogue Gravity Models.

In a particular  framework, in Analogue Gravity approach, Rivelles
\cite{riv} has shown  that NC $U(1)$ gauge theory, obtained by
Seiberg-Witten map \cite{sw} can induce an effective curved
spacetime. (For general reviews on NC quantum field theories and applications of
Seiberg-Witten map
see for example \cite{review}.) The effective metric is composed of the gauge
fields
and the anti-symmetric NC parameter $\theta _{\mu\nu}$. In a
similar vein it is very tempting to suggest that the Hamiltonian
(\ref{ham}) can be expressed as $H\sim
G^{\mu\nu}(F,J)\pi_\mu\pi_\nu $ with $G^{\mu\nu}\sim \eta
^{\mu\nu}+e(g-2)/m^2F^{\mu\beta } J_\beta ^{~\nu }$ as a simple
possibility. This identification was suggested  earlier  by \cite{jan} and by
some of us \cite{herve}. Then it is very revealing to compare the
works of \cite{riv} and ours. We find the  spin tensor $s_{\mu\nu
}$ in the spinning particle model can be identified with the NC
parameter $\theta _{\mu\nu}$ in \cite{riv}. In this way
noncommutativity and the resulting Analogue Gravity is not put by
hand from outside as is done conventionally (eg. \cite{riv}) but
appears as an effect of the particle spin. But for this to
materialise truely there is still a lot more work to do primarily
for two reasons: \\
(a) The NC parameter $\theta _{\mu\nu}$ is a $c$-number parameter
in \cite{riv} whereas $s_{\mu\nu}$ in the present case is a
dynamical variable. Hence some approximations
are needed.\\
(b) Interpreting $G^{\mu\nu}$ along the lines of Rivelles
\cite{riv} has to be done with care since the spacetime (or phase
space) is noncommutative in nature. It would have been more
convenient if there was a Darboux-like map that expresses the NC
variables in terms of canonical  variables (see for example
\cite{pal} for application of this approach in a different model).
Then the NC or spin induced effects will appear as additional
interaction terms but everything expressed in terms of canonical
variables. In fact this is not unlike the Seiberg-Witten map
\cite{sw}, at least in spirit. Work is in progress along these
lines.  \vskip .4cm {\it{Acknowledgements}}: SG wishes to thank
Professor Masud Chaichian for helpful discussions. He is also
grateful to the Physics Department, University of Helsinki,
Finland, where part of this work was done and to the Regular
Associateship Programme, ICTP, for providing the financial
support. Lastly we are grateful to the Referee for the constructive comments.

\end{document}